\documentclass[journal]{IEEEtran}

\usepackage{blindtext}
\usepackage{graphicx}
\usepackage{amsmath}
\usepackage{mathpazo}	

\usepackage{tikz}
\usepackage{color}
\usetikzlibrary{scopes,matrix,positioning}
\tikzset{
  mymx/.style={matrix of math nodes,nodes=myball,column sep=4em,row sep=-1ex},
  myball/.style={draw,circle,inner sep=0pt},
  mylabel/.style={midway,sloped,fill=white,inner sep=0pt,outer sep=1pt,below,
    execute at begin node={$\scriptstyle},execute at end node={$}},
  plain/.style={draw=none,fill=none},
  sel/.append style={fill=green!10},
  prevsel/.append style={fill=myblue1!40},
  route/.style={-latex,thick},
  selroute/.style={route,myblue1}
}

\usepackage{booktabs} 
\usepackage{colortbl} 
\usepackage{multirow}
\usepackage{xfrac}
\usetikzlibrary{patterns}
\usepackage{caption}	
\usepackage{subcaption}	

\usepackage{algorithm}
\usepackage[noend]{algpseudocode}
\usepackage{newcent}
\usepackage{lipsum}  

\usepackage{graphicx}

\definecolor{myblue}{RGB}{78,101,148} 
\definecolor{myblue1}{RGB}{61,66,98} 
\definecolor{myred}{RGB}{216,41,0}
\definecolor{myred_soft}{RGB}{216,115,86}
\definecolor{mygrey}{RGB}{128,128,128}
\definecolor{lightGray}{RGB}{220,220,220}

\ifCLASSINFOpdf
\else

\fi
\newcommand{\fjv}[1]{{\textcolor{black}{#1}}}

\begin{document}
\title{In-band Perturbation based OSNR Estimation}

\author{
	F.J. Vaquero-Caballero, D. Charlton, M.E. Mousa-Pasandi, D.J. Ives, C. Laperle,\\
	M. Hubbard, M. Reimer, M. O'Sullivan, S.J. Savory
\thanks{FJVC (e-mail: fjv24@cam.ac.uk), DJI, and SJS are with the Electrical Engineering Division, Department of Engineering, University of Cambridge, Cambridge CB3 0FA, UK.}	
\thanks{DC, MP, CL, MH, MR, and MOS, are with Ciena Corporation, Ottawa, Ontario K2K 0L1, Canada}
\thanks{Manuscript received April 19, 2005; revised August 26, 2015.}}

\markboth{Journal of \LaTeX\ Class Files,~Vol.~14, No.~8, August~2015}%
{Shell \MakeLowercase{\textit{et al.}}: Linear and Nonlinear Noise Separation in Optical Communications}

\maketitle

\begin{abstract}
OSNR is a figure of merit that measures the linear noise introduced in an optical link. OSNR is commonly measured by interpolating the linear noise from out-of-band measurements, requiring guard-bands. Current Dense Wavelength Division Multiplexing (DWDM) systems with higher spectral efficiency and tighter channel separation are particularly challenging for measuring OSNR. In-band methods are being considered to resolve this limitation. The OSNR of a single channel amplified link can be estimated by introducing a set of perturbations in a transmitted signal. Nonlinear noise is perturbation dependent while ASE noise is independent of the introduced perturbation. That axiom is exploited for the separation of optical noise. A least mean square fitting is used to estimate OSNR with a root mean square error of \fjv{0.2} dB.
\end{abstract}


\IEEEpeerreviewmaketitle

\section{Introduction}
Present DSP assisted optical transmission based on optical electric field modulation relies on the synthesis of a controlled transmit optical electric field.
To this end an optical modulator, capable of the transduction from drive instructions to the real and imaginary optical electric field on one or two polarizations, implements the desired field. We show this degree of control of the transmitted spectrum provides a means of partitioning noise contributions present in the communication channel. In particular, we demonstrate by modelling the estimation of Optical Signal-to-Noise Ratio (OSNR) from a set of transmitted spectra based solely on measurements of the intra-channel field. In comparison with other reported spectral methods\cite{Gariepy2018} the intra-channel nature of this method makes the noise estimate tolerant to optical filters present in most networks.

The method is compatible with spectral detection using the receiver of a transceiver or an optical spectrum analyzer (OSA). The method can be extended to the estimation of intra- and inter-channel nonlinear noise as well as frequency dependent transceiver implementation noise. 

OSNR is a measure of signal power to Amplified Spontaneous Emission (ASE) noise power in a 0.1 nm reference bandwidth. ASE power spectral density (PSD) is fairly constant in the vicinity of an optical channel. OSNR can be easily measured with an OSA. Operator demands for high spectral efficiency leads to reduction of guard-bands between optical channels, required for standard OSNR measurement which interpolates an estimate of ASE from the out of band portion of the optical channel. 

It follows that intra channel methods of OSNR estimation have been pursued. Reported methods exploit temporal correlations\cite{Shiner2020,Caballero18}, constellation shapes\cite{CaballeroVaquero2018}, repetitive patterns\cite{Cai2019}, and properties of the spectrum\cite{Gariepy2018}. In some cases, machine learning has also been considered for the extraction of OSNR\cite{CaballeroVaquero2018,Cho2020,Tanimura2019a,Lin2018,Lonardi2019a}. \fjv{Intrusive approaches such as polarization nulling\cite{Lee2006} or on-off signals have also been studied. Such approaches modify the power of the system, potentially disrupting the optical line system and affecting the measured results.}

In this work, we perturb the transmitted spectrum, thereby modifying the nonlinear noise contribution after propagation to ultimately estimate the OSNR. \fjv{The proposed approach maintains the power of the signal constant, avoiding power dynamics.} 

\fjv{Similar perturbations have been considered in wireless communications and transmitter calibration \cite{Measurement2011}, but to the best of our knowledge, this is the first time that this approach is considered in OSNR estimation for optical communications.}
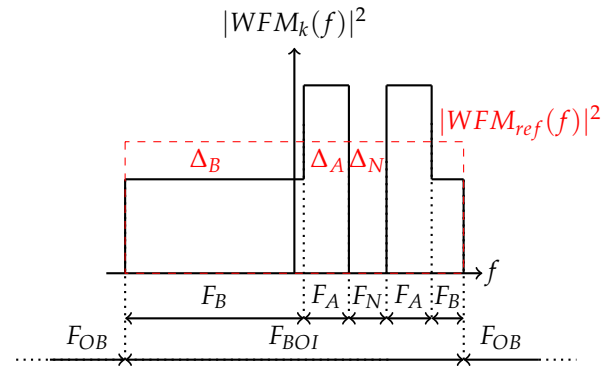
\begin{figure}[!h]
\centering
\begin{tikzpicture}
\newcommand\WFMfppnp[3]{
%
\pgfmathsetmacro{\PPLpointR}{0.15}
\pgfmathsetmacro{\PPLpointL}{0.75}
\pgfmathsetmacro{\PPRpointL}{0.15}
\pgfmathsetmacro{\PPRpointR}{0.45}  
\pgfmathsetmacro{\WFMwidth}{2.25}   %
\pgfmathsetmacro{\AG}{1}   %
\pgfmathsetmacro{\BG}{-0.5}   %

\draw [->,thick](-\WFMwidth-0.25+#1,#2)	            -- (\WFMwidth+0.25,#2)node[right=-0.1] {$f$};	
\draw [thick]   (-\WFMwidth+#1,#2)		            -- (-\WFMwidth+#1,1.75+\BG+#2);			    	    
\draw [->,thick](#1,#2)    			                -- (#1,3+#2) node[above] {#3}; 			        
\draw [thick]   (\WFMwidth+#1,#2)                   -- (\WFMwidth+#1,1.75+\BG+#2);
\draw [thick] 	(-\WFMwidth+#1,1.75+\BG+#2)         -- (0.875+#1-\PPLpointL,1.75+#2+\BG);      
\draw [thick]	(0.875+#1-\PPLpointL,1.75+\BG+#2)   --	(0.875+#1-\PPLpointL,1.5+#2+\AG); 
\draw [thick]	(0.875+#1-\PPLpointR,1.5+#2+\AG)    --	(0.875+#1-\PPLpointR,#2);	      
\draw [thick]	(0.875+#1-\PPLpointL,1.5+#2+\AG)    --	(0.875+#1-\PPLpointR,1.5+#2+\AG); 
\draw [thick]	(1.375+#1-\PPRpointL,1.5+#2+\AG)	--	(1.375+#1-\PPRpointL,#2);                   
\draw [thick]	(1.375+#1+\PPRpointR,1.5+#2+\AG)    --	(1.375+#1+\PPRpointR,1.75+\BG+#2);          
\draw [thick]	(1.375+#1-\PPRpointL,1.5+#2+\AG)    --	(1.375+#1+\PPRpointR,1.5+#2+\AG);           
\draw [thick] 	        (1.375+#1+\PPRpointR,1.75+\BG+#2)   --  (\WFMwidth+#1,1.75+\BG+#2);         
\draw [thick,<->]       (-\WFMwidth+#1,-0.675+0.125+#2-0.6) --	node[above] {$F_{BOI}$} (\WFMwidth+#1,-0.675+0.125+#2-0.6);
\draw [thick,  dotted]	(0.875+#1-\PPLpointL,#2-0.6)        --	(0.875+#1-\PPLpointL,1.5+#2+0.5);   
\draw [thick, dotted]	(0.875+#1-\PPLpointR,#2-0.6)        --	(0.875+#1-\PPLpointR,1.5+#2);       
\draw [thick, dotted]	(1.375+#1-\PPRpointL,#2-0.6)        --	(1.375+#1-\PPRpointL,1.5+#2);       
\draw [thick, dotted]	(1.375+#1+\PPRpointR,1.5+#2+0.5)    --	(1.375+#1+\PPRpointR,#2-0.6);       
\draw [thick, dotted]	(-\WFMwidth+#1,-0.675+0.125+#2-0.6) --	(-\WFMwidth+#1,1.75+\BG+#2);
\draw [thick, dotted]	(\WFMwidth2+#1,-0.675+0.125+#2-0.6) --	(\WFMwidth+#1,1.75+\BG+#2);

\draw [thick,<->] 	    (-\WFMwidth+#1,0.15+#2-0.75)        --  node[above] {$F_B$}     (0.875+#1-\PPLpointL,0.15+#2-0.75);
\draw [thick,<->] 	    (0.875+#1-\PPLpointL,0.15+#2-0.75)  --  node[above] {$F_A$}     (0.875+#1-\PPLpointR,0.15+#2-0.75);
\draw [thick,<->] 		(0.875+#1-\PPLpointR,0.15+#2-0.75)  -- 	node[above] {$F_{N}$}	(1.375+#1-\PPRpointL,0.15+#2-0.75);
\draw [thick,<->] 		(1.375+#1-\PPRpointL,0.15+#2-0.75)	-- 	node[above] {$F_{A}$}	(1.375+#1+\PPRpointR,0.15+#2-0.75);
\draw [thick,<->] 		(1.375+#1+\PPRpointR,0.15+#2-0.75)  -- 	node[above] {$F_{B}$}   (\WFMwidth+#1,0.15+#2-0.75);

\pgfmathsetmacro{\MidPointA}{1.375+#1+\PPRpointR/2-\PPRpointL/2}; %

\node[red] at (-1.175,1.5+#2){$\Delta_B$};
\node[red] at (0.45+#1,1.5+#2){$\Delta_A$};
\node[red] at (0.97+#1,1.5+#2){$\Delta_N$};

\draw [dashed,red] (-\WFMwidth+#1,1.75+#2) rectangle (\WFMwidth+#1,#2);
\draw [thick,<-]    (\WFMwidth+#1,-1.15+#2)     --  node[above] {$F_{OB}$} (\WFMwidth+#1+1,-1.15+#2);
\draw [thick,dotted](\WFMwidth+#1+1,-1.15+#2)   --  (\WFMwidth+#1+1+0.5,-1.15+#2);

\draw [thick,<-]    (-\WFMwidth+#1,-1.15+#2)    --  node[above] {$F_{OB}$} (-\WFMwidth+#1-1,-1.15+#2);
\draw [thick,dotted](\WFMwidth+#1-1,-1.15+#2)   --  (-\WFMwidth+#1-1-0.5,-1.15+#2);
}
\node[red] at (3,2) {$|WFM_{ref}(f)|^2$};
\WFMfppnp{0}{0}{$|WFM_k(f)|^2$};
\end{tikzpicture}
\caption{\centering Illustration of a Perturbed Spectrum with its corresponding frequency ranges}
\label{fig:PertConf}
\end{figure}

\begin{figure*}[h!]
\centering
\subfloat[\centering Evolution of normalized $P_{RX,k}^{(N)}$ for 2 dBm of launch power in absence of ASE.]{\includegraphics[scale=0.65]{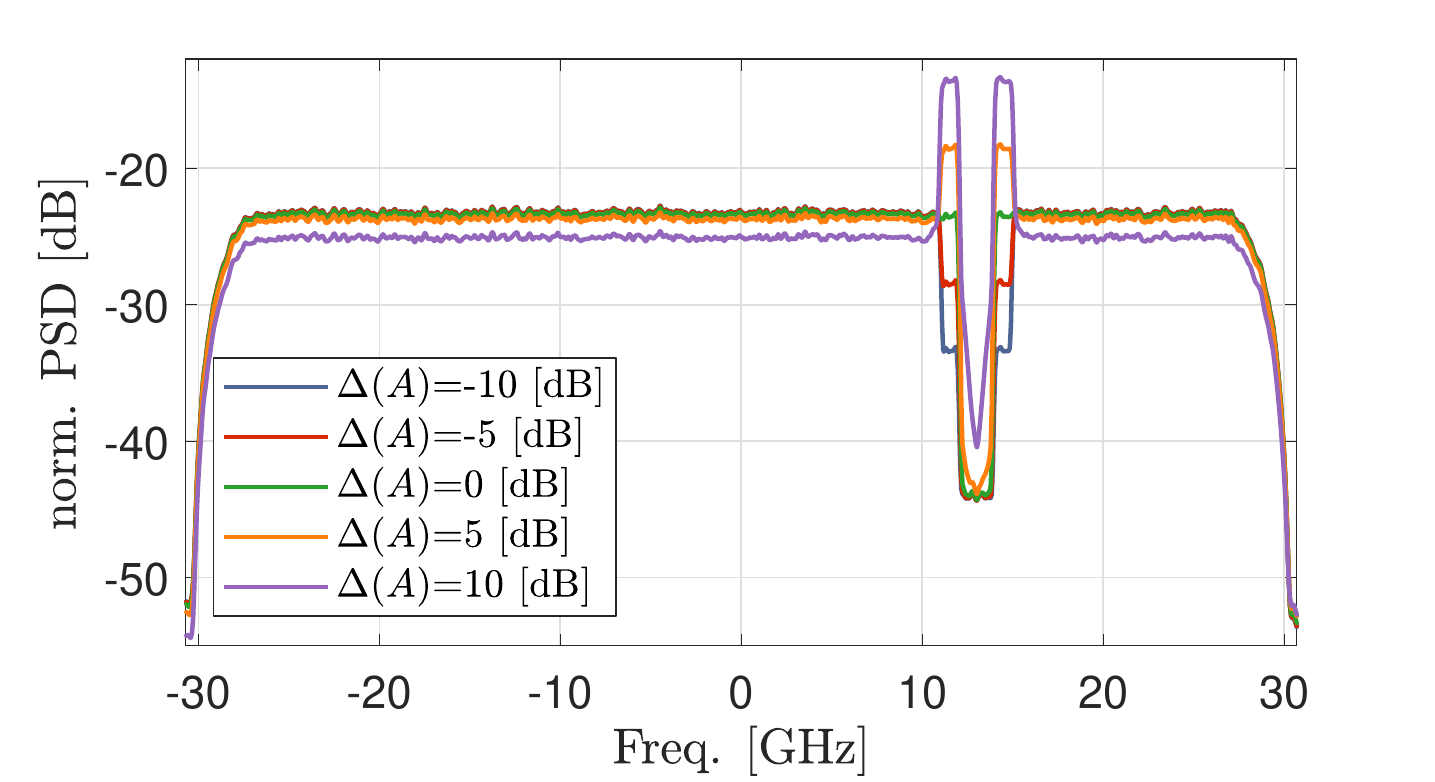}}
\subfloat[\centering Received spectra after 30 spans at 2 dBm launch  power and in absence of ASE.]{\includegraphics[scale=0.65]{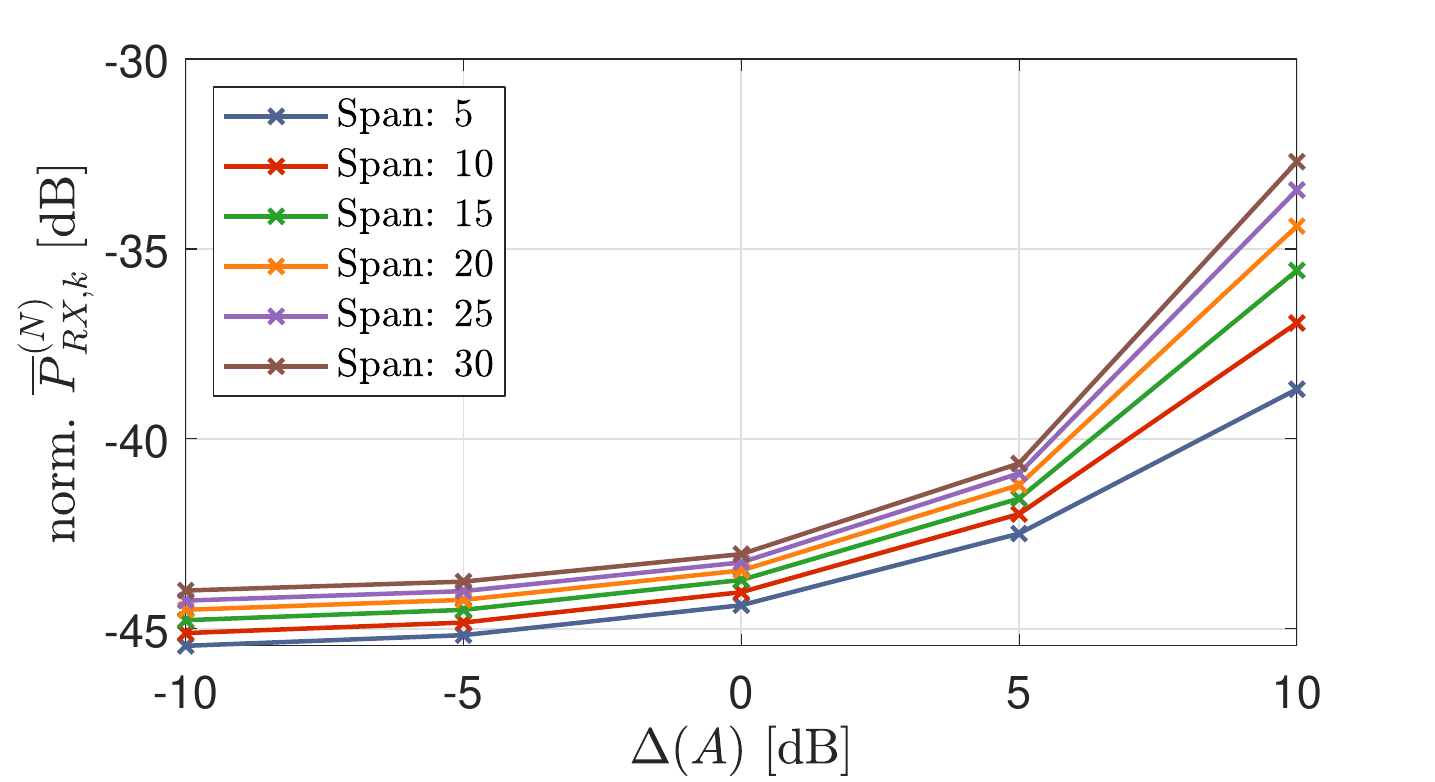}}
\caption{Spectral measurements of the perturbed spectrum for conditions of Table \ref{tab:simParameters}.}
\label{fig:SimulationMeasurement}
\end{figure*}
\section{Model Description}


Fig. \ref{fig:PertConf} illustrates a transmitted waveform instruction PSD $|WFM_k(f)|^2$, and the reference PSD $|WFM_{ref}(f)|^2$, which can be synthesized by an optical transmitter. Three spectral regions are defined: $F_A$, $F_B$ and $F_N$. We refer to the bandwidth of the spectrum as the bandwidth of interest (BOI) and its frequency extent: $F_{BOI}$. For completeness in Fig. \ref{fig:PertConf} we identify the out-of-band portion of the spectrum, i.e. the rest of the spectrum allocated to the other channels, as $F_{OB}$. Transmitted waveform PSDs $|WFM_k(f)|^2$ are written in terms of their piecewise values:
\begin{equation}
\begin{split}
|WFM_k(f)|^2	=	
\\
 \begin{cases}
 |WFM_{ref}(f)|^2 \Delta_k(A), \quad f \in F_A,\\
 |WFM_{ref}(f)|^2 \Delta_k(B), \quad f \in F_B,\\
 |WFM_{ref}(f)|^2 \Delta_k(N), \quad f \in F_N,\\
 \end{cases}
\end{split}
\end{equation}

$\Delta$ is defined in the linear domain as the ratio between the perturbed and the reference PSDs in a specific frequency range. For convenience, we refer to its logarithmic value as: $\Delta \text{ [dB]} = 10 \log_{10}(\Delta)$.

A transmission noise aggregation model consists of one (of many) transmitted waveform instruction PSDs ($|WFM_k(f)|^2$), transmitter noise PSD ($|NFL_{TX}(f)|^2$) (from  converters, DSP, analog RF and E/O), ASE noise PSD ($|ASE(f)|^2$), and propagation nonlinearity, approximated as an additive noise PSD \fjv{\cite{Carena2010}}, $|NLN_k(f)|^2$. Thus, assuming independence between the different PSD terms, the received PSD is:
\begin{equation}
\begin{split}
|RX_k(f)|^2	= |WFM_k(f)|^2 + |NFL_{TX}(f)|^2 \\+ |NLN_k(f)|^2 +	|ASE(f)|^2,
\end{split}
\end{equation}
where: \fjv{$|TX_k(f)|^2	= |WFM_k(f)|^2 + |NFL_{TX}(f)|^2$}. Fig. \ref{fig:SimulationMeasurement} (a) illustrates an RX spectrum for different values of $\Delta(A)$. Additionally, $|NFL_{TX}(f)|^2$ is assumed to be constant. This holds provided the peak-to-rms ratios of transmitted spectra used are similar.

\begin{figure*}[ht]
\centering
\subfloat[\centering Simple metric to compare the unpertubed NLN SNR in the absence of noise.]{\includegraphics[scale=0.65]{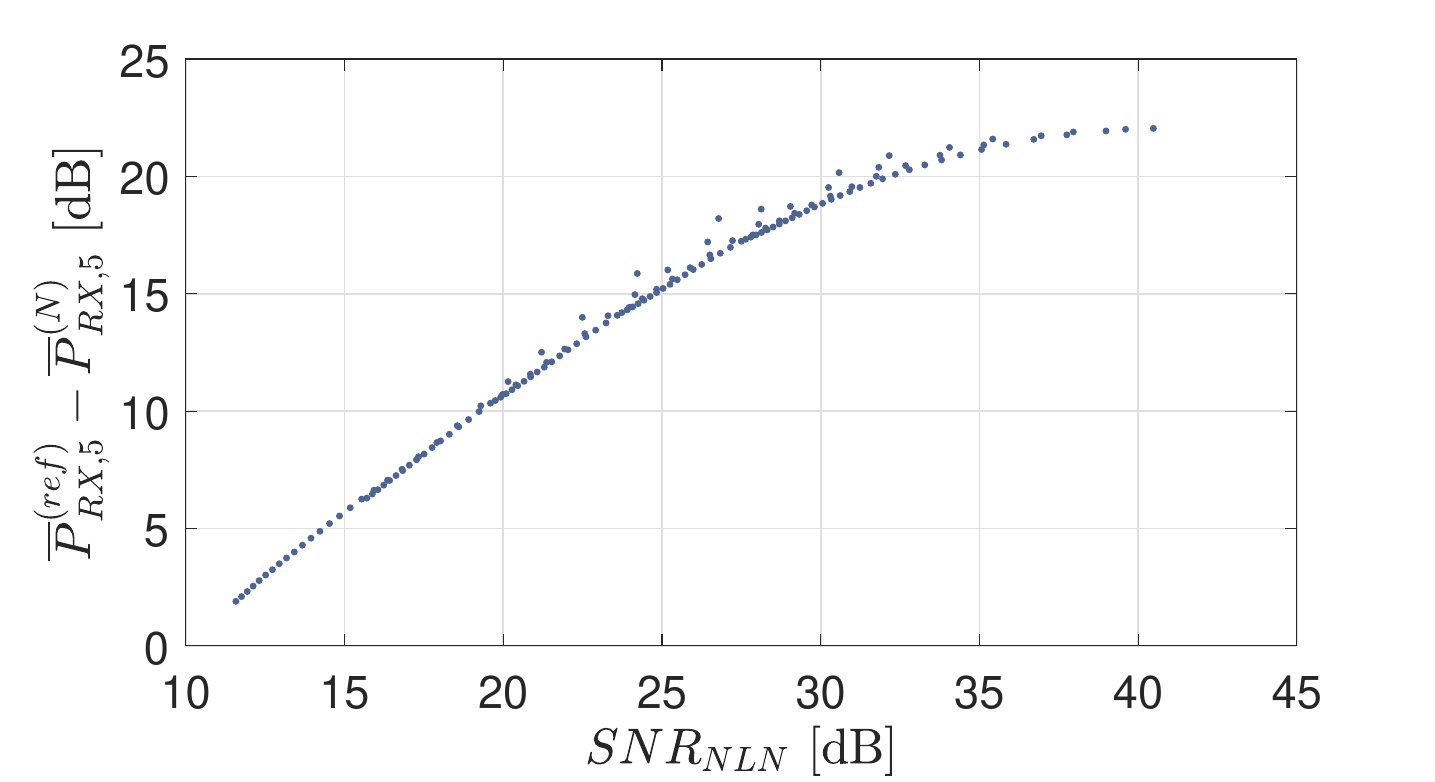}}
\subfloat[\centering Evolution of the fitting for the scenarios considered, OSNR in 0.1 nm.]{\includegraphics[scale=0.65]{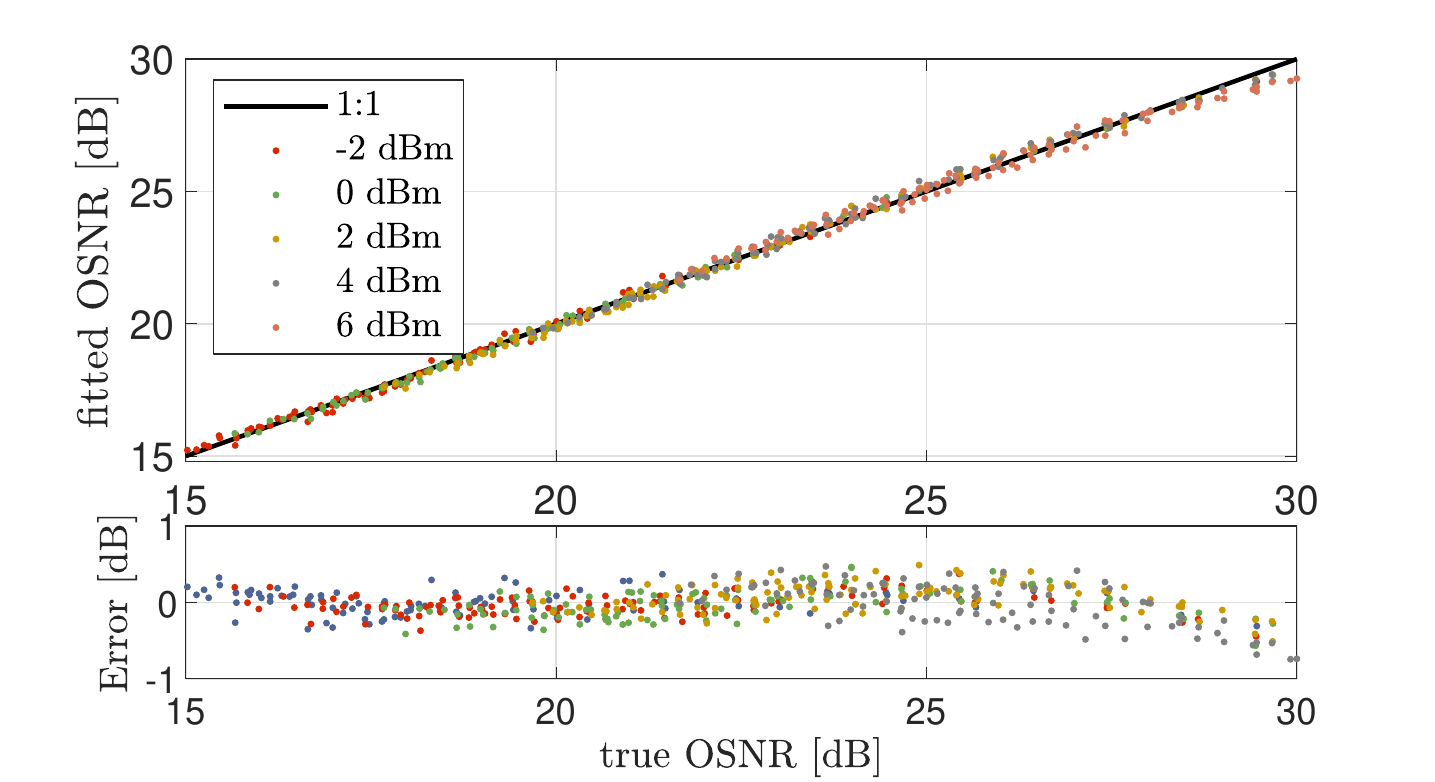}}  
\caption{Examples of metrics and results of the fitting for conditions of Table \ref{tab:simParameters}.}
\label{fig:EvoAndResults}
\end{figure*}
For this demonstration $\Delta_k(N)=0$, hence $\Delta_k(A)$ and $\Delta_k(B)$ can be related as:
\begin{equation}
\Delta_k(B) 	= \frac{1-K_{A} \Delta_k(A)}{K_B},
\label{eq:ConstPowerDef}
\end{equation}	
where $K_A$ and $K_B$ are the relative contributions in power of each region of the reference waveform:
\begin{equation}
\begin{split}
K_A =  \frac{\int_{f\in F_A} |TX_{ref}(f)|^2 df}{\int_{f \in F_{BOI}} |TX_{ref}(f)|^2 df}, \\
K_B =  \frac{\int_{f\in F_B} |TX_{ref}(f)|^2 df}{\int_{f \in F_{BOI}} |TX_{ref}(f)|^2 df},
\end{split}
\end{equation}
\fjv{Thus, we ensure the total power of the transmitted remains constant regardless of the applied perturbation.}

We consider an amplifier control (i.e. \fjv{constant} gain) such that the  ASE noise PSD is independent of the loaded waveform, while the NLN is frequency and perturbation dependent. Hence, it is possible to define two average PSD (APSD) integrated terms $\overline{P}_{RX,k}^{(ref)}$ and $\overline{P}_{RX,k}^{(N)}$:
\begin{equation}
\overline{P}_{RX,k}^{(ref)}	=	\int_{F_{ref}}	|RX_k(f)|^2	\,	df	/	\int_{F_{ref}}	\,	df,
\end{equation}

\begin{equation}
\overline{P}_{RX,k}^{(N)}		=	\int_{F_N}	|RX_k(f)|^2 \,	df		/	\int_{F_N}	\,	df,
\end{equation}
where $F_{ref}= F_A \cup F_B$. 
The evolution of APSDs as a function of $\Delta_k(A)$ is used to differentiate between noise components. \fjv{Spectra for the 30 spans case are shown in Fig.  \ref{fig:SimulationMeasurement} (a) for links parameterized by the coefficients of Table \ref{tab:simParameters}, while Fig. \ref{fig:SimulationMeasurement} (b) illustrates how the APSD in $F_N$ is enhanced by NLN as a function of $\Delta(A)$ and span count}.

\fjv{The transmission conditions of the simulation is summarized in Table \ref{tab:simParameters}}. A Dual Polarization Quadrature Phase Shift Keying (DP-QPSK) 56.8 GBaud Tx signal similar to a commercial product such as Ciena WaveLogic Ai was simulated with a notch at positive baseband frequency. The centers of the $F_A$ regions are at 11.5 and 14.5 GHz with a width of 1 GHz each, while $F_N$ is centered at 13 GHz with a width of 2 GHz. Only 80\% of $F_N$ is integrated, discarding the edges of the spectral region. Up to 30 spans of Non-Dispersion-Shifted Fiber (NDSF) were modeled. A 150 MHz OSA resolution OSA was used in simulations \fjv{to emulate a Finisar WaveAnalyzer 1500S} \fjv{with a Super-Gaussian filter response. The optimum launch power is around 3 dBm. The noise floor PSD contribution $|NFL_{TX}(f)|^2$ is chosen to be 22.5 [dB] below the PSD of an unperturbed transmitted WFM.} 
\fjv{The perturbation topology was chosen based on several criteria: band-guards are required to perform the integration in $F_N$ because of OSA resolution, and high concentration of power in $F_A$ must be avoided since it can result in driver nonlinearities.}

\begin{table}
   \centering
   \small
\caption{Parameters for the Simulation Verification.} \label{tab:Parameters}
\begin{tabular}{l | r  r r}
 \centering%
Parameter &	\multicolumn{2}{c}{Value} \\
 \hline	
Fiber Type							&	\multicolumn{2}{c}{NDSF, (G.652)}	\\
\hline
Symbol Rate (B) 	[Gbaud]		        &	\multicolumn{2}{c}{56.8}            \\
\hline
\fjv{Modulation format}             &	\multicolumn{2}{c}{\fjv{DP-QPSK}}   \\
\hline
\fjv{Sampling Frequency	[GHz]}	    &	\multicolumn{2}{c}{\fjv{170.4}}	\\
\hline
\fjv{number of symbols}	            &	\multicolumn{2}{c}{\fjv{$2^{17}$}}	\\
\hline
\fjv{Split-step step size [km]}	    &	\multicolumn{2}{c}{\fjv{0.01}}	\\
\hline
Nonlinear Coeff. ($\gamma$) [1/W/km]&	\multicolumn{2}{c}{1.3}		\\
\hline
Number of WDM Channels	 			&	\multicolumn{2}{c}{1}		\\
\hline
Dispersion Param. (D) [ps/nm/km]	&	\multicolumn{2}{c}{16.7}	\\
\hline
Attenuation ($\alpha$) [dB/km]		&	\multicolumn{2}{c}{0.2}		\\
\hline
Channel Launch Power [dBm]          &	\multicolumn{2}{c}{-2:2:\fjv{6}}	\\
\hline
Modulation Format					& 	\multicolumn{2}{c}{QPSK}	\\
 \hline
Span Length [km]					&	\multicolumn{2}{c}{100}		\\
 \hline
Number of Spans						&	\multicolumn{2}{c}{1:30}	\\
\hline
$\Delta_k(A)$	[dB]				&	\multicolumn{2}{c}{-10:5:10}\\
\hline
Root-Raised Cosine	Roll-Off		&	\multicolumn{2}{c}{0.07}	\\
\hline
\% $F_N$ Integrated					&	\multicolumn{2}{c}{80\%}	\\
\hline
$|NFL_{TX}(f)|^2$  [dB]		        &	\multicolumn{2}{c}{-22.5}	\\
\hline
Amplifier NF. [dB]					&	\multicolumn{2}{c}{4.5:1:7.5}\\
\hline
OSA Resolution [MHz]			    &	\multicolumn{2}{c}{150}\\
\hline
\end{tabular}
\label{tab:simParameters}
\end{table}%

\section{Estimation Methodology and Results}
\fjv{Fig. \ref{fig:EvoAndResults} (a) plots the relationship between the NLN SNR, $SNR_{NLN}$, of a transmitted unperturbed signal, and the logarithmic subtraction of $\overline{P}_{RX,5}^{(ref)}- \overline{P}_{RX,5}^{(N)}$. Such different of APSD is ultimately a power ratio between the APSDs in $F_{ref}$ and $F_N$ for $\Delta(A)=10$. The transmission scenarios are described on Table \ref{tab:simParameters}. } \fjv{Thus, a power ratio of two APSD in the notch region is a descriptive metric of NLN.}

\fjv{In this analysis, we propose to estimate OSNR based on different $\Delta(A)$, considering different ASE loadings scenarios. Hence, the estimation of OSNR becomes more intricate and requires of functions to translate the measured $\Delta(A)$ APSD transductions into OSNRs.}

For this demonstration, a least square fitting (LS) over several APSD terms is performed to estimate the OSNR. \fjv{LS was chosen given its simplicity and performance for OSNR estimation, more complex functions such as Neural Networks can also be considered}. The terms considered for the fitting are: 
\begin{itemize}
\item 	Constant term, ($k_0$)
\item 	$\overline{P}_{RX,k}^{(ref)}$ [dB], $\Delta(A)$	= -10 [dB], ($k_1$)
\item	$\overline{P}_{RX,k}^{(N)}$ [dB], $\Delta(A)$ 	=	-10:5:10 [dB], ($k_2$, $k_3$, $k_4$, $k_5$, $k_6$)
\end{itemize}
where $k_i$ terms refers to fitting coefficients multiplying each APSD\fjv{, and -10:5:10 nomenclature means between -10 to 10 in steps of 5}. Hence, OSNR is estimated as:
\begin{equation}
OSNR \, [dB] = k_0 + k_1 \overline{P}_{RX,k}^{(ref)} + k_2 \overline{P}_{RX,1}^{(N)} + \cdots +  k_6 \overline{P}_{RX,5}^{(N)},
\end{equation}
\fjv{where the evolution of NLN in $F_N$ as a function of $\Delta(A)$ is studied and compared with the APSD in $F_{ref}$ as baseline.} Only one $\overline{P}_{RX,k}^{(ref)}$ term was included since \fjv{APSD TX contribution in $F_{ref}$} is constant regardless of the perturbation \fjv{(by Equation \ref{eq:ConstPowerDef})}, and the only contribution which is $\Delta(A)$-dependent is the NLN. The NLN variation with $\Delta(A)$ over $\overline{P}_{RX,k}^{(ref)}$ can be \fjv{considered negligible}.

Fig. \ref{fig:EvoAndResults} (b) illustrates the results of the fitting \fjv{constrained to OSNR$\leq$30 dB with standard deviation error of 0.2 dB}. \fjv{Different colors illustrate the error of the different launch powers, confirming the validity of the methodology and the homogeneity of the estimation error in the linear and the nonlinear regime}.

\section{On the Potential for In-Service Measurements}
\fjv{The introduction of perturbations in a transmitted signal can result in significant penalties, but transmitted signal can be engineered to include perturbations with minimal penalties. Coding can be applied to the transmitted signal to obtain the desired spectral shape with minimal penalties. Moreover, the spectrum can be expanded to accommodate a dummy signal with the desired perturbation while maintaining the information signal intact.}

\begin{figure}
    \centering
    \includegraphics[scale=0.65]{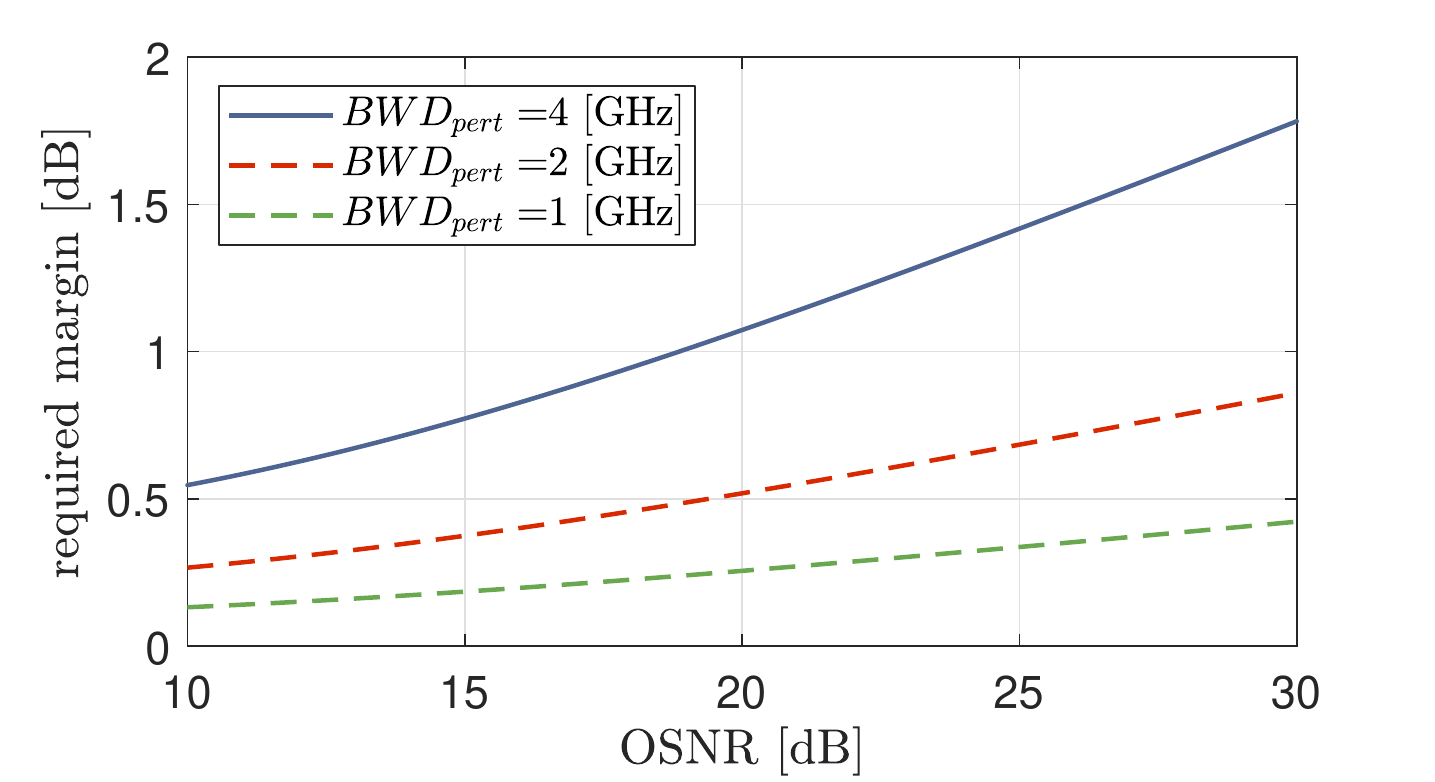}
    \caption{Penalty as a function of the unperturbed SNR.}
    \label{fig:MarginFunctionPert}
\end{figure}

\fjv{ If the total bandwidth is kept constant, part of it can be allocated to perturbations. Given that minimization of NLN requires of 2-6 Gbaud sub-carriers \cite{Poggiolini2016a}, a single sub-carrier can be scarified. Figure \ref{fig:MarginFunctionPert} illustrates the OSNR margin required to maintain capacity for different perturbation bandwidths $BWD_{pert}$, where the perturbed SNR ($SNR'$) based on Shannon \cite{Shannon48}, its capacity can be calculated as:} 
\begin{equation}
SNR' = (1+SNR)^{\frac{B}{B-BWD_{pert}}}-1,
\end{equation}
\fjv{where $B$ is the baud-rate of the transmitted signal. Smaller perturbation bandwidths are included for reference, requiring smaller OSA bandwidths.}
\section{Conclusions}
\fjv{In this paper, we adapt the wireless application of notch to optical communications, combining a notch with positive perturbations to estimate OSNR in the presence of nonlinearities.}

Nonlinear noise is estimated from a set of different transmitted spectra by means of an LS fitting to allow an estimate of OSNR. A standard deviation error of 0.16 dB was obtained for OSNRs lower than 30 dB.

Future work will explore additional methods for noise separation based on perturbations with special interest in analytical approaches, and their experimental verification. \fjv{Coherent measurements may require smaller band-guards in $F_N$, and it may result in smaller operational penalties}. Machine learning approaches with potentially more flexibility and higher accuracy than the LS discussed here will also be considered.

\section*{Acknowledgment}
FJVC would like to thank Ciena for his industrial PhD sponsorship, and for insightful team discussions which have significantly improve the content and the quality of this paper.

\ifCLASSOPTIONcaptionsoff
  \newpage
\fi

\bibliographystyle{ieeetr}
\bibliography{library}

\
\end{document}